# A Study of Recent Contributions on Information Extraction

Parisa Naderi Golshan
M.Sc. Student
Electrical and Computer Department
University of Zanjan
Zanjan, Iran
Email: p.naderigolshan@znu.ac.ir

HosseinAli Rahmani Dashti
M.Sc. Student
Electrical and Computer Department
University of Zanjan
Zanjan, Iran
Email: srahmani@znu.ac.ir

Shahrzad Azizi
M.Sc. Student
Electrical and Computer Department
University of Zanjan
Zanjan, Iran
Email: azizi.shahrzad@znu.ac.ir

Leila Safari
Assistant Professor
Electrical and Computer Department
University of Zanjan
Zanjan, Iran
Email: lsafari@znu.ac.ir

*Abstract*— **This paper reports on modern approaches in Information Extraction (IE) and its two main sub-tasks of Named Entity Recognition (NER) and Relation Extraction (RE). Basic concepts and the most recent approaches in this area are reviewed, which mainly include Machine Learning (ML) based approaches and the more recent trend to Deep Learning (DL) based methods.**

## I. INTRODUCTION

Today, a significant portion of textual information, including online news, government documents, military texts, legal acts, medical records, and court sentences, are provided through unstructured free text documents. Seeking information from these free text form documents is a challenging task that requires advanced analytical approaches and innovative methods for knowledge discovery. Rule based methods and advanced ML based methods have attracted a number of researchers for IE and knowledge discovery in many areas mainly news and medical domain. Neural networks have also been used in this area. More recently, the DL based approaches have been the focus of interest of some researcher. All of these approaches have led to the emergence of IE technologies [1]. The extracted information usually is represented in a structured form to be used in other natural language related tasks like Question Answering (QA) or for enhancement of current knowledge bases like *DBpedia* [2].

IE is one of the most prominent text-mining techniques, which aims to analyze unstructured texts to identify information, or events that are explicitly or implicitly expressed in the text [3, 4]. The task of IE is to identify a set of predefined concepts within a given domain and ignore the unrelated information. A domain consist of a corpus comes with information that is clearly specified. In other words, IE aims at extracting real information constructed from unstructured text and converting it to the structured text [1, 3]. The process of extracting information involves identifying some small structures such as phrases, referring to an individual, group, or geographic references, and also numerical expressions, and finally finding meaningful relationships between them [1]. In this scenario, the domain knowledge is required to accurately collect the extracted partial information in a structured way.

This paper is organized as follows: Section II briefly covers the basic definitions and relevant concepts to IE and its sub-tasks. Section III presents the summary of main methods that have been used for IE in recent years, mainly ML based and knowledge based methods followed by a comparison of the reviewed methods and their evaluation. Finally, sections IV and V explain the conclusion and comparisons.

## II. CONCEPTS

IE methods have encountered a rapid growth during the past years. These methods can be divided into two main categories including Knowledge Engineering (KE) and ML based methods. KE methods include knowledge-based and rule-based approaches that mainly used to integrate knowledge in computer systems to solve problems requiring human expertise, as well as in Data Mining (DM) systems for the production of rules [5]. The second category, ML based approaches use learning techniques. In contrast to KE based methods, ML based methods do not require rules for manual extraction, but they need to know the scope and functionalities of the system. The ML techniques are divided into supervised, semi-supervised and unsupervised classes [5].

In addition, as a sub-category of ML based methods, DL is used in a variety of domains and applications including Computer Vision, Speech Recognition, Natural Language Processing, Social Network Filtering, Bioinformatics, and Drug Design. A review of some recent works [6, 7, 8, 9, 10] revealed that these methods are suitable in solving complex learning problems and more specifically they have represented successes in the NLP field.

The task of IE involves identifying instances of entities, relationships, and events together with extracting relationships between them. The extracted information then is provided in the form of a defined template or object by the user, each contains the attributes that are discovered by IE. The IE main tasks include NER, Co-reference Resolution (CO), RE and Event Extraction (EE) [1, 11]:

a) *NER*: Detection and classification of types of entities, such as organization, person and location, and also temporal, numerical, and currency expressions.

b) *CO*: Finding all the terms that refer to an entity.

c) *RE*: Identifying the relationship between entities in the text. For example, the relationship between a person and a place.

d) *EE*: The goal is to identify everything that has happened.

In this review, the authors have only focused on the two essential sub-tasks of NER and RE. A brief description of these two subtasks and their related concepts are provided later in this section, while more details of the reviewed approaches in NER and RE is presented in section III.

NER involves identifying references to certain types of objects, such as names of individuals, companies, and places [12]. Additionally, a significant amount of relevant information may be provided implicitly in the text, which requires referring to prior or external knowledge to provide a full understanding of the provided information in the natural language or free form text.

ML based methods have been extensively used for NER in many languages mainly English, German, Spanish, Dutch, Japanese, Chinese, French, and so on. Learning methods for NER include supervised, semi-supervised, and unsupervised learning. Supervised techniques including the Hidden Markov Model (HMM) [13], Decision Tree (DT) [14], Maximum Entropy Model (MEM) [15, 16], Support Vector Machine (SVM) [17], and Conditional Random Field (CRF) have been used in many NER tasks in various domains [18]. Semi-supervised (Semi-monitored) methods use both labeled and unlabeled corpus. The well-known semi-supervised method is bootstrapping, starts with a small set of initial seeds, and generates and stores more annotations at each time and repeats these steps to reach a certain threshold. To be able to use these methods for NER, a large number of features and annotations required to be specified. However, most languages suffer from the lack of large annotated corpus. So, some unsupervised techniques were introduced to address this issue. Clustering is the well-known unsupervised method, which is used for NER in [12, 19].

RE is the other important sub-task of IE that aims to identify the relationships between entities in the text and represents them as structured information. For example, the relationship Works-For can be extracted from the text for entities such as Person and Organization, which relates a person and an organization together [20].

Due to the capabilities of the DL methods, there is a recent trend in using them in most Text Mining and Natural Language Processing (NLP) applications. Generally, most DL methods and more specifically the recently proposed approaches for RE have been established based on some common concepts [8].

Word Embedding and Positional Embedding are the most important concepts, which are extensively used in the area of DL based NLP tasks. Word Embedding is collective name for language modeling and feature learning techniques in NLP, in which words or phrases are mapped from a vocabulary to a real number vector [8]. As the words that are close to an entity may have more useful information, the relative distance of each word with the corresponding entity is determined using Positional Embedding [8]. In addition, Convolutional Neural Network (CNN) and Recurrent Neural Networks (RNN) are common across most DL models for RE:

CNN is one of the major DL methods, which can be trained in several layers in a powerful way. This method is efficient and common in different computer vision applications. In general, a CNN network consists of three main layers of the convolutional layer, a pooling layer, and a fully connected layer, each of which performs various tasks [21].

RNN was actually designed to process sequential signals. In an ordinary neural network, all inputs and outputs are independent of one another, but in many cases, this assumption is not valid. The most special feature of RNN is the hidden state that stores the information of a sequence. Also, we do not necessarily need to have an output at any one time or have an entry. RNNs are called recursive because the output of each layer depends on the calculations of the previous layers. In other words, these networks have a memory that stores information about the data being viewed [22]. Some other systems are also proposed based on the combination of neural network models [9].

III. METHODS

Several innovative approaches are proposed for IE in recent years, in which some of them have been reviewed and investigated in current work.

*A. Named Entity Recognition*

NER is used in many applications such as DM and QA applications to search for the information about the People or Organization Names. The web includes unstructured information such as news articles, scientific articles, blogs and topic archives, and email communications and so on. NLP claims that it can extract such unstructured information that is hidden from the machine and represents it in a structured form. The proposed algorithms analyze some elements that occur in the sentence and identify named entities such as Person or Organization Names, Locations and Time. Then, they categorize these entities to increase the reusability of the extracted information [23].

Reference [24], has presented a system that uses Multi-Layer Perceptron (MLP) method on the Clinical TempEval corpus to identify event spans. The words, part-of-speech tags (POS-tags), and shape information are the features that have been used for training in this system. The system firstly runs the classifier to identify the event spans and then identifies the attribute values.

Reference [25], proposes an approach that consists of three main steps:

   *a) Standard data set acquisition:* In this research, the Iranian Student News Agency (ISNA) is used as the source of the news, which is a reliable source among Persian corpora. It uses an automated searching and a crawling procedure is used in this work to download ISNA site content. It starts from the ISNA home page and stores all web pages in a database. Then apply a preparation and filtering phrase is applied on them and finally, an XML file is created for each NEWS of the corpus.

   *b) Preprocessing and Normalization:* This step involves a series of processes that apply to the raw data to adapt them to the intended requirements to reduce the ambiguity of the language and improve the precision of the algorithm. Tokenization, Filtering Stop Words, Case Transforming, Filtering Tokens by length, and Making N-gram are actions that are performed at this stage.

   *c) Qualification and Verification of Validness:* The purpose of this step is to evaluate the validity and reliability of the categories in the dataset.

In [26] three approaches for NER, called Baseline Method, Bootstrapping, and Card Pyramid Parsing are investigated. Here is a brief summary of each one.

1. *Baseline Method:* This method tries to solve two constraints in rule-based and supervised-based systems. Human intervention does not require for labeling the training data. In addition, the system can examine more than three types of classical entities.

2. *Bootstrapping:* This method considers the part of the tagged data as seed, labels them as example data, and the adds to the training set.

3. *Card Pyramid Parsing:* A card pyramid is a tree-like graph with a root, internal nodes, and leaf nodes so that with n leaves we have exactly n levels. It is worth noting that, since many nodes have two parents, the card pyramid is not a tree. In this method, initially, all the entities in the sentence are placed as pyramid leaves in the same order as in the text. The leaf label is a type of relevant entity. If the two entities are not connected, then their label will be Non-Relation. Otherwise, their associated type is tagged. This will continue until reaching the root of the pyramid [20].

Several ML based methods have been used to identify biomedical entities, which have achieved good results on GENIA corpus. In [10], neural networks are used on untagged biomedical text files to generate potential information and represent them as word vectors. More specifically, the proposed method of [10] is a Biomedical Named Entity Recognition method (Bio-NER) based on a deep neural network architecture with several layers. Each layer summarizes its features based on the features produced by the lower layers.

Reference [27], IE techniques, patterns were handcrafted or semi-supervised learned have been used to auto-mark. Here, ML is used to learn patterns or learn some of the basics of KE. The steps are as the follows: 1. Learning step. 2. Translate the patterns learned in step 1, as needed for the marking process. 3. Filtering for possible patterns. 4. Test on educational data. 5. Test on test data.

*B. Relation Extraction*

Many applications in IE and IR need to understand the semantic relationships between entities like persons [28].

Reference [29] has introduced a system for extracting information from films based on Wikipedia encyclopedia information and resources published on the Web. The proposed method is based on the Bootstrapping that attempts to answer the Semantic Drift problem and provide an algorithm called Improved Pattern Ranking Algorithm (IPRA). At each step, firstly, the sentences containing the input sample are extracted from the related articles and transformed into the desired and defined model. Secondly, using a confidence calculation, the algorithm checks whether the pattern is extracted from a good sample or not. Then the extracted patterns are ranked and converted to the samples and finally, the samples have been used for evaluation of other samples.

Content Analyzer and IE System (CAINES) [30], is an experimental system to extract knowledge and security information from electronic documents. This system is based on the KE approach and uses sublanguage analysis techniques. It can extract information from reports of terrorist incidents. The system uses lexicons with syntactic and semantic structures instead of using statistics alone. With regard to the architecture, the preprocessing unit of CAINES initially processes the news texts and sends raw information to the conceptual equivalence unit in order to integrate the expressions. After the creation of the conceptual structure of the news, the pattern-matching unit of CAINES eliminates the sentences that do not have useful information, taking into account the constraints of the extraction patterns. Thus, the accepted sentences are passed to activate the output template unit and, the relevant information is extracted. Some sentences represent an event and can activate the template while the others can only complete the news. Therefore, if necessary, the corresponding template is activated. After selecting the output template, the sentences are processed to fill the template, and finally, the filled template is displayed in the output.

Reference [31], an attempt to move from traditional approaches to approaches that are more modern. It introduces a CNN for extracting a relationship that automatically learns the features of sentences and minimizes the dependence of external NLP supervised resources to the features. It uses multiple

TABLE I. EVALUATION OF THE REVIWED SYSTEMS

| System | Main Task | Year | Method | Data | Measurement (%) | | |
|---|---|---|---|---|---|---|---|
| | | | | | *P* | *R* | *F1* |
| [27] | **NER** | 2005 | **HMM (ML)** | Penn Treebank Corpus | Not available | | |
| Persica[25] | **NER** | 2012 | **KNN, NB, SVM** | ISNA | Acc = 68.03 | | |
| [10] | **NER** | 2015 | **Bio-NER** | GENIA Corpus | 66.54 | 76.13 | 71.01 |
| [24] | **NER** | 2017 | **CNN** | Clinical TempEval corpus | 78.8 | 78.8 | 78.8 |
| MV-RNN [22] | **RE** | 2012 | **RNN** | Movie Reviews (IMDB) | Acc = 79.5 | | |
| CAINES [30] | **RE** | 2015 | **Knowledge Engineering** | Reports on Terrorist Incidents | 92 | 89 | 90.48 |
| [31] | **RE** | 2015 | **CNN** | ACE 2005 | 71.25 | 53.91 | 61.3 |
| IExM [29] | **RE** | 2017 | **Distant-Supervised Algorithm** | Movie Articles | 80.1 | 69.4 | 74.4 |
| Card Pyramid [20] | **NER&RE** | 2010 | **Rule-Based** | TREC corpus | 94.2 | 92.1 | 93.2 |
| TIE [33] | **NER&RE** | 2010 | **Temporal Entropy** | TimeBank | Acc = 69.5 | | |
| [35] | **NER&RE** | 2016 | **CNN** | i2b2-2010 | 76.34 | 67.35 | 71.2 |

window sizes and position embedding for filtering and encoding relative distances, respectively.

While the traditional method of IE, suffers from causing errors in relationship detection, CNN approaches have proved to perform much better performance. The main reasons are their ability in automatically learning the characteristics of sentences, and also the less dependency on external tools and resources. In [31] present an IE system, which is created by the crude sentences determined by the position of the institutions. Only the words, n-grams, and their position in sentences are extracted as features in this work. It then uses the combination of word embedding that can detect latent semantic properties and a CNN that can recognize classes of n-gram (for classifying and modeling sentences). Four main layers of look-up tables, recognizing n-grams, pooling and logistic regression layer are used to extract the relationship in this work.

Moreover, some works have considered both NER and RE tasks:

Reference [32], presents an IE system for medical data in the Italian language. Due to the problems and costs of building and obtaining tagged data for non-English language, an unsupervised method has been used in this system which has two main steps:

In the first step, domain entities are extracted using a standard NLP tool. In the second step, it attempts to extract the relationships between each entity pair of the entire text. The proposed, unsupervised method attempts to perform auto-tagging.

The Temporal Information Extraction (TIE) is another IE system that is presented in [33]. It deals with extracting facts from the text and identifies the temporal relations between the extracted time spans and event. TIE processes each sentence in two steps: First, using a syntactic parser and a semantic role marker that generate a set of attributes, it extracts events and identifies time phrases. Second, it uses its probabilistic model to determine the inequality relations between the extracted endpoints.

Some other works have been focused on Extraction of Semantic Relation to be used in QA systems, IR systems, Ontology Learning (OL), and Semantic Web Annotation (SWA). In [34] the context pattern has been used to semantically derive relationships in a few steps. First, the sentence framework pattern is obtained by lexical analysis. Then the syntax tree model which is obtained via a syntactic analysis calculates the weight of the words using the syntax tree pattern. Finally, semantic relationships are extracted using the Bootstrap Semi-Supervised ML Method.

Reference [35], proposed a method for RE in the biomedical domain. It focused on using engineered features or kernel methods to create a feature vector. The features are then fed to a classifier to predict the correct class [36]. Clinical discharge summaries are used in this work for RE between clinical entities. The main purpose is exploiting the power of the CNN to automatically learn the features and reduce the dependency to the manual feature engineering process.

IV. COMPARISION AND EVALUATION METRICS

All of the reviewed systems evaluated using the standard metrics of Accuracy, Precision (P), Recall (R) and F-Measure.

These metrics are based on Confusion Matrix with four parameters, True Positive (TP), False Positive (FP), True Negative (TN), and False Negative (FN). The Accuracy is the rate of the true detection in the system result. Precision is the fraction of retrieved documents that are relevant to the query and Recall is the fraction of the relevant documents that successfully retrieved.

$$\text{Accuracy} = \frac{TP + TN}{TP + TN + FP + FN} \quad (1)$$

$$\text{Precision} = \frac{\text{\# of relevant retrived docs}}{\text{\# of retrived docs}} \quad (2)$$

$$\text{Recall} = \frac{\text{\# of relevant retrived docs}}{\text{\# of relevant docs}} \quad (3)$$

$$F1 = 2 * \frac{\text{Precision} * \text{Recall}}{\text{Precision} + \text{Recall}} \quad (4)$$

A summary of some of the reviewed systems and their evaluations are provided in the Table I. The main task of each system which is NER, RE or both NER and RE are shown in the table as well. As it is clear from the table, among the four systems provided for NER, the most recent one that has used CNN performed the best with the F1 measure of 78.8 per cent, however the performance metric for the HMM based NER (the first in the table) is not available. Also, the three ML based methods used on ISNA corpus only showed the accuracy of 68.03.

Among the proposed methods for RE in the table, the CAINES system has used the KE based method and performed the best with the F1 measure of more than 90 per cent. In addition, the proposed solution for RE in [31] has the second highest F1 measure of 82.8 percent.

Moreover, the three solutions of Table I proposed for both NER and RE. The Rule-based system which is applied on the TREC corpus [20], have shown the highest F1 measure of 93.2. However, the generated rules of the method are very specific to the system and it is very difficult to be generalized. The other two systems [32, 33], have shown the performance measures of 71.2 of F1 measure on i2b2-2010 corpus and accuracy of 69.5 on TimeBank corpus respectively.

It is important to notify that, the data and the evaluation metrics used in the proposed methods are different. Therefore, it is not possible to provide an exact comparison of them. Also, some of the proposed methods only aimed to provide a solution for NER, some others tried to deal with RE issues, while the others aimed to provide solutions for both NER and RE. So, it is not possible to compare all the solution together. We tried to compare them roughly to extract any trends might have been found in the proposed approaches. To be able to provide an exact decision on feasibility and performance of the proposed solution it is required to use similar bases and data. Finally, the current review reveals a recent trend to develop these systems based on DL and CNN methods.

V. CONCLUSION

A review of several innovative approaches in the area of IE is provided in current work, which aims to exploit the capability of the approaches for NER and RE as the two main sub-tasks of the IE. Based on our investigation, the ML based approaches are popular for both NER and RE tasks. However, due the lack of annotated data in most languages and domains, some unsupervised and semi-supervised methods have been used for IE and its sub-tasks as well. In addition, rule based and KE based methods have shown a good performance in some tasks, however, they suffer from the lack of generalizability and dependency on specific domains and knowledge resources. Moreover, we encounter a trend to DL based methods in recent years, which aim to reduce the dependency to the external resources and knowledge bases and try to learn from the features of data to provide more generalizable solutions for IE.